\documentclass[]{spie}  
\usepackage{acronym}
\usepackage[]{graphicx}

\usepackage{textcomp}
\newcommand{\pb}{\textsc{Polarbear}}

\newcommand{\sa}{Simons Array}

\newcommand{\uKrts}{$\mu\mbox{K}\sqrt{\mbox{s}}$}
\newcommand{\rtHz}{$\sqrt{\mbox{Hz}}$}

\acrodef{tes}[TES]{Transition Edge Sensor}
\acrodef{snr}[SNR]{Signal-to-Noise Ratio}
\acrodef{hwp}[HWP]{Half-Wave Plate}
\acrodef{sq}[SQUID]{Superconducting Quantum Interference Device}
\acrodef{cmb}[CMB]{Cosmic Microwave Background}
\acrodef{fts}[FTS]{Fourier Transform Spectrometer}
\acrodef{pei}[PEI]{polyetherimid}
\acrodef{uhmwpe}[UHMWPE]{ultra-high molecular weight polyethylene}
\acrodef{ar}[AR]{Anti-Reflection}
\acrodef{lsn}[LSN]{Low-stress silicon nitride}
\acrodef{htt}[HTT]{Huan Tran Telescope}

\title{Development and characterization of the readout system for POLARBEAR-2} 

\author{
D. Barron\supit{k},
P.A.R. Ade\supit{x},
Y. Akiba\supit{aa},
A.E. Anthony\supit{b,e},
K. Arnold\supit{k},
M. Atlas\supit{k},
A. Bender\supit{v},
D. Boettger\supit{k},
J. Borrill\supit{c,z},
S. Chapman\supit{g},
Y. Chinone\supit{m,j},
A. Cukierman\supit{j},
M. Dobbs\supit{v},
T. Elleflot\supit{k},
J. Errard\supit{z,c},
G. Fabbian\supit{a,o},
C. Feng\supit{k},
A. Gilbert\supit{v},
N. Goeckner-Wald\supit{j},
N.W. Halverson\supit{b,e,l},
M. Hasegawa\supit{m,aa},
K. Hattori\supit{m},
M. Hazumi\supit{m,aa,q},
W.L. Holzapfel\supit{j},
Y. Hori\supit{m},
Y. Inoue\supit{aa},
G.C. Jaehnig\supit{b,l},
A.H. Jaffe\supit{h},
N. Katayama\supit{q},
B. Keating\supit{k},
Z. Kermish\supit{i},
R. Keskitalo\supit{c},
T. Kisner\supit{c,z},
M. Le Jeune\supit{a},
A.T. Lee\supit{j,w},
E.M. Leitch\supit{d,p},
E. Linder\supit{w},
F. Matsuda\supit{k},
T. Matsumura\supit{n},
X. Meng\supit{j},
N.J. Miller\supit{t},
H. Morii\supit{m},
M.J. Myers\supit{j},
M. Navaroli\supit{k},
H. Nishino\supit{q},
T. Okamura\supit{m},
H. Paar\supit{k},
J. Peloton\supit{a},
D. Poletti\supit{a},
C. Raum\supit{j},
G. Rebeiz\supit{f},
C.L. Reichardt\supit{y},
P.L. Richards\supit{j},
C. Ross\supit{g},
K.M. Rotermund\supit{g},
D.E. Schenck\supit{b,e},
B.D. Sherwin\supit{j,r},
I. Shirley\supit{j},
M. Sholl\supit{w}, 
P. Siritanasak\supit{k},
G. Smecher\supit{ab},
B. Steinbach\supit{j},
N. Stebor\supit{k},
R. Stompor\supit{a},
A. Suzuki\supit{j},
J. Suzuki\supit{m},
S. Takada\supit{s},
S. Takakura\supit{u,m},
T. Tomaru\supit{m},
B. Wilson\supit{k},
A. Yadav\supit{k},
O. Zahn\supit{w}
\skiplinehalf
\supit{a}AstroParticule et Cosmologie, Univ Paris Diderot, CNRS/IN2P3, CEA/Irfu, Obs de Paris, Sorbonne Paris Cit\'e, France \\
\supit{b}Center for Astrophysics and Space Astronomy, University of Colorado, Boulder, CO 80309 \\
\supit{c}Computational Cosmology Center, Lawrence Berkeley National Laboratory, Berkeley, CA 94720 \\
\supit{d}Department of Astronomy and Astrophysics, University of Chicago, Chicago, IL 60637 \\
\supit{e}Department of Astrophysical and Planetary Sciences, University of Colorado, Boulder, CO 80309 \\
\supit{f}Department of Electrical and Computer Engineering, University of California, San Diego, CA 92093 \\
\supit{g}Department of Physics and Atmospheric Science, Dalhousie University, Halifax, NS, B3H 4R2, Canada \\
\supit{h}Department of Physics, Imperial College London, London SW7 2AZ, United Kingdom \\
\supit{i}Department of Physics, Princeton University, Princeton, NJ 08544 \\
\supit{j}Department of Physics, University of California, Berkeley, CA 94720 \\
\supit{k}Department of Physics, University of California, San Diego, CA 92093-0424 \\
\supit{l}Department of Physics, University of Colorado, Boulder, CO 80309 \\
\supit{m}High Energy Accelerator Research Organization (KEK), Tsukuba, Ibaraki 305-0801, Japan \\
\supit{n}Institute of Space and Astronautical Science (ISAS), Japan Aerospace Exploration Agency (JAXA), Sagamihara, Kanagawa 252-5210, Japan \\
\supit{o}International School for Advanced Studies (SISSA), Trieste 34014, Italy \\
\supit{p}Kavli Institute for Cosmological Physics, University of Chicago, Chicago, IL 60637 \\
\supit{q}Kavli Institute for the Physics and Mathematics of the Universe (WPI), Todai Institutes for Advanced Study, The University of Tokyo, Kashiwa, Chiba 277-8583, Japan \\
\supit{r}Miller Institute for Basic Research in Science, University of California, Berkeley, CA 94720 \\
\supit{s}National Institute for Fusion Science, Toki, Japan \\
\supit{t}Observational Cosmology Laboratory, Code 665, NASA Goddard Space Flight Center, Greenbelt, MD 20771 \\
\supit{u}Osaka University, Toyonaka, Osaka 560-0043, Japan \\
\supit{v}Physics Department, McGill University, Montreal, QC H3A 0G4, Canada \\
\supit{w}Physics Division, Lawrence Berkeley National Laboratory, Berkeley, CA 94720 \\
\supit{x}School of Physics and Astronomy, Cardiff University, Cardiff CF10 3XQ, United Kingdom \\
\supit{y}School of Physics, University of Melbourne, Parkville, VIC 3010, Australia \\
\supit{z}Space Sciences Laboratory, University of California, Berkeley, CA 94720 \\
\supit{aa}The Graduate University for Advanced Studies, Hayama, Miura District, Kanagawa 240-0115, Japan \\
\supit{ab}Three-Speed Logic, Inc., Vancouver, B.C., V6A 2J8, Canada 
}

\authorinfo{Contact Author: Darcy Barron (drbarron@ucsd.edu)}

\begin{document} 
\maketitle 

\begin{abstract}
\pb-2 is a next-generation receiver for precision measurements of the polarization of the cosmic microwave background (\ac{cmb}). Scheduled to deploy in early 2015, it will observe alongside the existing \pb-1 receiver, on a new telescope in the \sa on Cerro Toco in the Atacama desert of Chile. For increased sensitivity, it will feature a larger area focal plane, with a total of 7,588 polarization sensitive antenna-coupled \ac{tes} bolometers, with a design sensitivity of 4.1 \uKrts. The focal plane will be cooled to 250 milliKelvin, and the bolometers will be read-out with $40\times$ frequency domain multiplexing, with 36 optical bolometers on a single SQUID amplifier, along with 2 dark bolometers and 2 calibration resistors. To increase the multiplexing factor from $8\times$ for \pb-1 to $40\times$ for \pb-2 requires additional bandwidth for SQUID readout and well-defined frequency channel spacing. Extending to these higher frequencies requires new components and design for the LC filters which define channel spacing. The LC filters are cold resonant circuits with an inductor and capacitor in series with each bolometer, and stray inductance in the wiring and equivalent series resistance from the capacitors can affect bolometer operation. We present results from characterizing these new readout components. Integration of the readout system is being done first on a small scale, to ensure that the readout system does not affect bolometer sensitivity or stability, and to validate the overall system before expansion into the full receiver. We present the status of readout integration, and the initial results and status of components for the full array.
\end{abstract}


\acresetall
\section{Introduction}
\label{sec:intro}
The \pb\ experiment is measuring fluctuations in the polarization of the cosmic microwave background (CMB), with the goal of characterizing the gravitational lensing signal at small angular scales, as well as the signal from inflationary gravitational waves at large angular scales. The first gravitational lensing results from \pb-1 were recently published\cite{PB_CLdd_2014,PB_GalaxyCross_2014,PB_CLbb_2014}. 

The \pb\ experiment is located at the Ax Observatory on Cerro Toco, at an altitude of 5200 meters in the Atacama desert of Chile. The observing site is a dry, high altitude location, minimizing the emission and absorption of atmospheric water vapor and oxygen at millimeter wavelengths. The \pb-1 receiver\cite{Kermish_SPIE2012,Arnold_SPIE2012,Barron_LTD2014} started observations in 2012, installed on the 3.5 meter Huan Tran Telescope (HTT). The Simons Array expansion will add two additional telescopes, identical to HTT, which are currently under construction. The \pb-2 receiver is a new receiver with improved sensitivity that will deploy onto the first of these new telescopes in 2015, observing alongside \pb-1.

In this paper, we will discuss the the readout system and the requirements placed on it from the overall instrument design and specifications. Further details can be found in the accompanying papers in this proceedings on the Simons Array\cite{Arnold_SPIE2014}, \pb-2\cite{Hazumi_SPIE2014}, detector and cold filter development\cite{Hattori_SPIE2014}, and the room temperature electronics\cite{Bender_SPIE2014}.

\section{Instrument Design} 
\label{sec:design}
Improving the sensitivity of CMB measurements requires expanding to many more detectors, since the sensitivity of transition-edge sensor (TES) bolometers observing at millimeter-wave is limited by photon statistics\cite{Arnold2010}. Observations at multiple frequencies are also necessary, in order to separate out the underlying CMB signal from astrophysical foreground contamination. The \pb-1 focal plane has 1,274 detectors observing at a single frequency, 150 GHz. \pb-2 will have a larger focal plane area, with 1897 new dichroic pixels. Each pixel has four transition edge sensor (TES) bolometers, coupled to a sinuous antenna with two frequency bands at 95 GHz and 150 GHz, and two polarization orientations, so the focal plane will have a total of 7,588 detectors. After \pb-2 is deployed, a copy of this receiver will be made and installed on the third telescope of the Simons Array. The original \pb-1 receiver will finally be replaced with a third receiver with this style of expanded multichroic array. The full Simons Array of three telescopes will include 11,382 multichroic pixels with 22,764 detectors observing at three different frequencies.

\begin{table}[h]
\caption{Design specifications for the \pb-2 receiver} 
\label{tab:specs}
\begin{center}       
\begin{tabular}{|l|l|} 
\hline
\rule[-1ex]{0pt}{3.5ex}  Frequencies & 95 GHz and 150 GHz  \\
\hline
\rule[-1ex]{0pt}{3.5ex}  Number of pixels & 1897 pixels   \\
\hline
\rule[-1ex]{0pt}{3.5ex}  NET bolometer & 360 \uKrts \\
\hline
\rule[-1ex]{0pt}{3.5ex}  NET array & 4.1 \uKrts  \\
\hline
\rule[-1ex]{0pt}{3.5ex}  Detector temperature & 250 mK  \\
\hline
\rule[-1ex]{0pt}{3.5ex}  Field of view & 4.8 degrees  \\
\hline
\rule[-1ex]{0pt}{3.5ex}  Beam size & 5.2 arcmin at 95 GHz, 3.5 arcmin at 150 GHz \\
\hline 
\end{tabular}
\end{center}
\end{table}

\section{Readout system and requirements}
\label{sec:readout}

TES detectors are thermistors with a steep power-resistance relation at their superconducting transition. The TES detectors are operated with a low-impedance voltage bias, which causes electro-thermal feedback (ETF) that keeps the total power (optical power and electrical power) at a constant level. Incident optical power is then converted into a changing current, which can be measured by Superconducting Quantum Interference Devices (SQUIDs).
In the frequency-domain multiplexing (fMux) readout system, described in detail in Ref.~\citenum{Dobbs2012}, multiple detectors are read out continuously on the same set of wires by separating their signals in frequency space. Each detector in a module is voltage biased at a unique frequency, defined by a cold resonant filter with an inductor and capacitor in series with the detector. Current from the bolometers are summed and pre-amplified by series-arrays of  SQUIDs, cooled to 4 Kelvin. The signal is amplified and demodulated by room-temperature electronics. 

The increased number of detectors requires the multiplexing factor to increase as well, due constraints from thermal loading from wiring, space inside the cryostat, cost, and other factors. The multiplexing factor will increase from $8\times$ for POLARBEAR-1 to $40\times$ for POLARBEAR-2, and will eventually increase to $64\times$ for the Simons Array. This requires additional bandwidth for SQUID readout and well-defined frequency channel spacing. While expanding the readout system to this higher multiplexing factor, we must ensure that detector performance is not compromised by factors like readout noise, electrical crosstalk, and detector stability.

\subsection{SQUID amplifiers}
\label{sec:squids}

Current from the bolometers is summed and pre-amplified by SQUIDs at 4 Kelvin, coupled through a superconducting input coil. This coil's low impedance preserves the voltage bias across the bolometers, which have a resistance of approximately $1 \Omega$. Series-array SQUIDs can achieve a high transimpedance amplification, and the SQUID output voltage is read out with room-temperature amplifiers. 

\begin{figure}
   \begin{center}
   \begin{tabular}{c}
   \includegraphics[height=7cm]{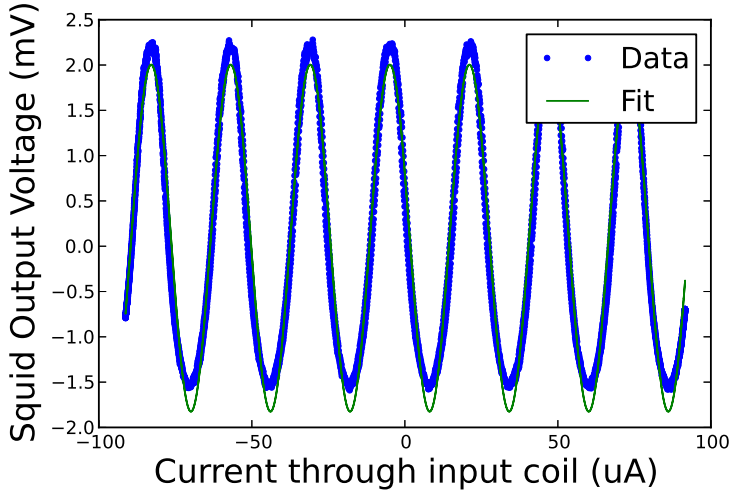}
   \end{tabular}
   \end{center}
   \caption[example] 
   { \label{fig:sqfit} 
Output response of a typical SQUID series array, with model to estimate transimpedance.}
   \end{figure}

\begin{figure}
   \begin{center}
   \begin{tabular}{c}
   \includegraphics[height=7cm]{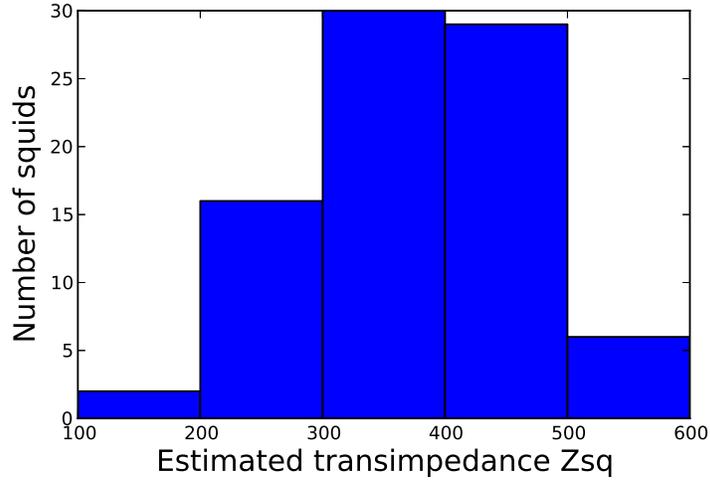}
   \end{tabular}
   \end{center}
   \caption[example] 
   { \label{fig:sqdist} 
Distribution of estimated transimpedance values for a wafer of SQUID series arrays.}
   \end{figure}

The response of a SQUID can be approximated as
	\begin{equation}
	\label{eq:sqout}
V^{SQ}_{OUT} = 1/2 V_{pp}\sin(2 \pi I_{IN}/I_{\phi_{0}}) , 
	\end{equation}
where $V^{SQ}_{OUT}$ is the output voltage, $I_{IN}$ is the current through the input coil, $V_{pp}$ is the peak-to-peak output voltage, and $I_{\phi_{0}}$ is the input current which produces a quantum of magnetic flux. The relation of input current to flux quanta is fixed by geometry, but the transimpedance depends on the SQUID temperature and the current bias applied. The operating point is set along the downward sloping linear region of this curve, where the transimpedance $Z_{SQ}$ is approximately $ \pi V_{pp} / I_{\phi_{0}}$. For our devices, manufactured at NIST\cite{Huber2001}, $V_{pp}$ ranges from $1.5 - 5$ mV, and $I_{\phi_{0}} \approx 25 \mu A$. The SQUIDs undergo initial screening where they are dunked in liquid helium and the output response is measured and $V_{pp}$ and the transimpedance is determined. An example showing the fit to the simple sine wave approximation is shown in Figure~\ref{fig:sqfit}. Results from screening a wafer of SQUIDs is shown in Figure~\ref{fig:sqdist}. SQUIDs with high transimpedance are selected and undergo further characterization to find optimal bias values and determine noise properties. 

The requirements for SQUID performance are calculated based on keeping the SQUID readout noise contributions subdominant to the bolometer power noise terms. This power noise is referred to the input of the SQUID as a current noise in order to compare to readout noise. Each SQUID readout noise terms is expected to be subdominant to the bolometer power noise contribution, which is when the optical loading is the lowest. The first SQUID readout noise contribution is fundamental SQUID noise, which is around 3.5 pA\rtHz. The room-temperature amplifier also contributes noise, with a specification of 1.2 nV\rtHz\cite{Dobbs2012}. To keep this contribution low, below 3.1 pA\rtHz, there is a minimum requirement on the transimpedance of each SQUID, $Z_{SQ} > 400 \Omega$. 

We are investigating two different SQUID designs for use in \pb-2 and the future Simons Array receivers. The existing SQUID design used in \pb-1 is a 100 element series array of SQUIDs, with a typical array noise of 3.5 pA\rtHz, and a bandwidth of 120 MHz. With screening, these SQUIDs will have acceptable properties to keep their readout noise within specifications. However, there are still gains to be made with reducing this noise by reducing the fundamental SQUID noise and increasing the transimpedance, which would help improve the overall $NET_{array}$. The alternate SQUID design, also made by NIST, consists of banks of 64 element arrays, that can be connected in series and parallel combinations, for example a 3 by 2 array, depending on the desired configuration for output impedance and transimpedance. We are continuing to test different designs of SQUIDs for their noise and transimpedance properties, and will screen for the best elements to include in the final receivers.

For \pb-1 and other first-generation fMux systems, the SQUIDs were operated using a flux locked loop with shunt feedback\cite{Dobbs2012}. This shunt feedback works to keep the SQUID in the linear region of its output, extending the dynamic range of input current. However, shunt feedback has a limited usable bandwidth of about 1.3 MHz, and increasing the multiplexing factor will require extending the usable readout bandwidth to several MHz. For \pb-2 and other experiments with higher multiplexing factor, a new feedback scheme will be used which is called Digital Active Nulling (DAN), which is described in Ref.~\cite{deHaan2012}. With DAN, the current at the SQUID is nulled in a digital feedback loop for each bolometer frequency channel. The bandwidth of the new electronics with DAN implemented is now 10 MHz.

\begin{figure}
   \begin{center}
   \begin{tabular}{c}
   \includegraphics[height=7cm]{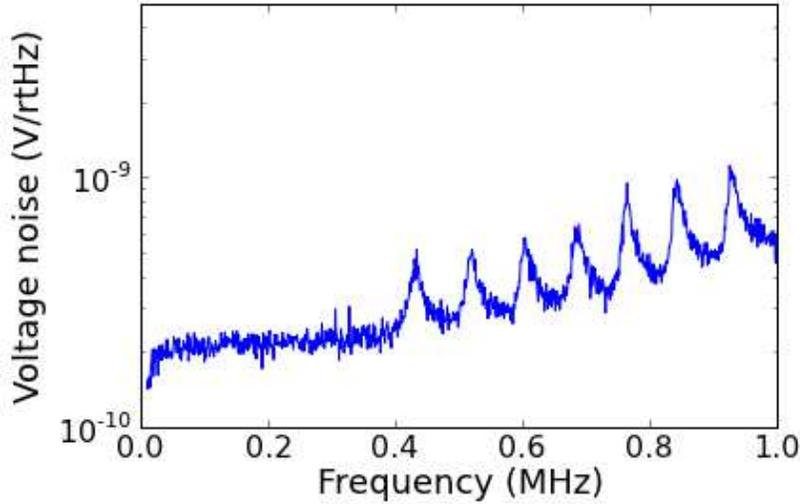}
   \end{tabular}
   \end{center}
   \caption[example] 
   { \label{fig:sqnoise} 
Voltage noise vs frequency for a SQUID in a flux-locked loop, with Johnson noise peaks from bolometers at 1.5 Kelvin.}
   \end{figure}

With increased readout bandwidth, the low pass RF filtering on the cryostat, achieved with pi filters, will need to be relaxed out to several MHz. Noise pickup from auxiliary electronics, telescope motors, and other sources will need to be characterized and eliminated from the readout band. Figure~\ref{fig:sqnoise} shows a typical noise spectral analysis for the SQUID output from the \pb-1 system, showing the noise floor with Johnson noise peaks at the bolometer frequencies. 
	
\subsection{Frequency schedule for multiplexing}
\label{sec:LC}

The readout channels are defined by an LC filter with an inductor and a capacitor in series with each bolometer. For \pb-1, these were 16 $\mu$H inductors made by NIST along with commercial ceramic capacitors. For \pb-2, the increase in readout bandwidth and decrease in channel spacing requires significant improvements in the inductor and capacitor properties. The equivalent series resistance (ESR) of commercial capacitors increases with frequency, and at the higher end of the readout bandwidth this would become so large that it is comparable to the bolometer resistance, affecting bolometer performance and stability\cite{Hattori2013}. Interdigitated capacitors with low ESR are now being developed and fabricated at UC Berkeley\cite{Hattori_SPIE2014}. Current limits to the physical size and capacitance of these components sets the lower end to the frequency band, 1.5 MHz. The ESR has been shown to be very low at the upper readout frequencies around 3 MHz. The inductance value will be increased to 60 $\mu$H, which makes the resonance peak narrower and allows closer channel spacing. These inductors and capacitors are being fabricated together on the same chip, and the fabrication process will allow us to achieve tight tolerances in values so that the final channel locations and spacing are well-constrained. The requirement on electrical crosstalk is that it is less than the optical crosstalk, less than 1 percent. The layout and spacing of the LC components will ensure that crosstalk from neighboring components is minimal. 

Stray inductance from wiring in series with the detectors has an impedance that rises with frequency, and must remain small compared to the $1\Omega$ bolometer impedance at the highest readout frequency. This subKelvin wiring is used to connect the SQUIDs at 4 Kelvin with the LC boards at 250 mK, and must be long enough to make these connections as well have minimal thermal loading on the cold stages. This wiring will be commercially made NbTi striplines on polyimide film, with low inductance and low thermal conductivity. The current requirement for the \pb-2 wiring is a maximum length of 60 cm, and maximum inductance of 45 nH. These constraints will keep the stray inductance at an acceptable level for our readout bandwidth and channel spacing.

\begin{figure}
   \begin{center}
   \begin{tabular}{c}
   \includegraphics[height=7cm]{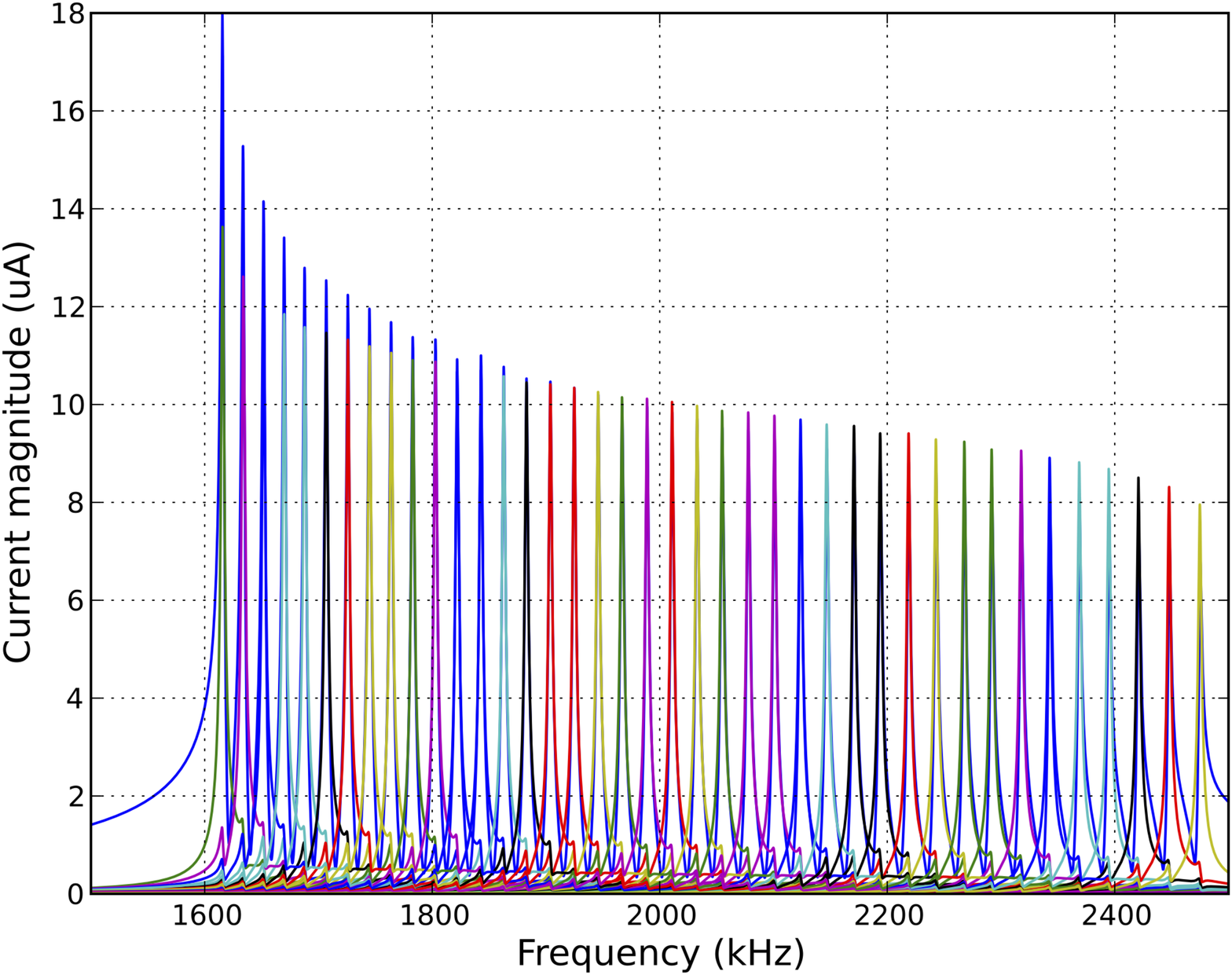}
   \end{tabular}
   \end{center}
   \caption[example] 
   { \label{fig:sim_comb} 
Simulated frequency response for a 40 channel frequency multiplexing setup, showing admittance of the resonant filters, each consisting of a 60 $\mu$H inductor and an interdigitated capacitor }
   \end{figure}

With these constrants for the lower and upper bounds for the readout bandwidth and the expected channel spacing, the result is that the currently achievable multiplexing factor is 40, shown in Fig.~\ref{fig:sim_comb}, with evenly spaced channels between 1.6 and 2.5 MHz. We expect to be able to expand the usable bandwidth and increase to $64\times$ multiplexing for the next receivers in the Simons Array.

\subsection{Detector array characterization}

The \pb-2 design is an array with an NET of 4.1 \uKrts\, based on 7,588 detectors with an $NET_{bolo}$ of 360 \uKrts\ . Many factors can reduce the final array sensitivity, including excess noise and yield. The end-to-end readout system, from warm electronics to cold components, will be validated before expanding to read out the entire array. The overall expected readout noise contribution is expected to be 7 pA\rtHz\ . Detailed screening and characterization of components, including the SQUIDs and LC filters described above, is being performed at several institutions to select components for the final receiver. The detector wafers are also tested and screened to check that they meet the expected bolometer properties like saturation power and frequency band placement. The detectors must also be characterized in the integrated system for characteristics like stability and sensitivity. This process will be ongoing as we continue after the \pb-2 receiver is complete to commission the second and third Simons Array receivers. 

The final array yield and sensitivity will be determined by observations on the sky, as described in Ref.~\cite{Kermish_SPIE2012}. During observations, the SQUIDs are tuned daily during a cryogenic cycle, and the bolometers are typically tuned every hour to adjust for changing atmospheric conditions. Robust software that can use saved detector properties to quickly re-tune the entire array of detectors, without any user interaction, is important for observation efficiency. The stability of the detectors to changing power is also important, since detectors that go unstable from atmospheric fluctuations must be shut off until the next daily cryogenic cycle.

\section{Conclusion}
\label{sec:conclusion}
We have presented the current status of the readout system for the \pb-2 experiment. Development and testing is ongoing at several institutions within the \pb\ collaboration. The full array and supporting readout electronics will be assembled and characterized in the laboratory at KEK in Japan, before deploying to the Simons Array in 2015. The full complement of Simons Array receivers is expected to be complete in 2016.
\acknowledgments

Calculations were performed on the Central Computing System, owned and operated by the Computing Research Center at KEK, and the National Energy Research Scientific Computing Center, which is supported by the Department of Energy under Contract No. DE-AC02-05CH11231. 
The \pb\ project is funded by the National Science Foundation under grants AST-0618398 and AST-1212230.
The \sa\ expansion of \pb\ is funded by The Simons Foundation. 
The KEK authors were supported by MEXT KAKENHI Grant Number 21111002, and acknowledge support from KEK Cryogenics Science Center. 
The McGill authors acknowledge funding from the Natural Sciences and Engineering Research Council of Canada, the Canada Research Chairs Program, and Canadian Institute for Advanced Research. 
BDS acknowledges support from the Miller Institute for Basic Research in Science, NM acknowledges support from the NASA Postdoctoral Program. 
MS gratefully acknowledges support from Joan and Irwin Jacobs.
All silicon wafer-based technology for \sa\ will be fabricated at the UC Berkeley Nanolab and the UCSD Nano3 Microfabrication Laboratory. 
We are indebted to our Chilean team members, Nolberto Oyarce and Jos\'e Cortes. 
The James Ax Observatory operates in the Parque Astron\'{o}mico Atacama in Northern Chile under the auspices of the Comisi\'{o}n Nacional de Investigaci\'{o}n Cient\'{i}fica y Tecnol\'{o}gica de Chile (CONICYT).


\bibliography{Barron_SPIE2014}
\bibliographystyle{spiebib}   

\end{document}